\documentclass[twocolumn,nofootinbib,amsmath,amssymb,aps,prd,balancelastpage,superscriptaddress]{revtex4-1}

\usepackage{color}
\usepackage[active]{srcltx}
\usepackage{amsmath,amsfonts,amssymb,amsthm,amstext,amscd,eucal,srcltx}
\usepackage{epsfig,graphicx,bm}
\usepackage{epstopdf, epsf}
\usepackage{dcolumn}
\usepackage{hyperref}

\newcommand{\be}{\begin{equation}}
\newcommand{\ee}{\end{equation}}

\newcommand{\bse}{\begin{subequations}}
\newcommand{\ese}{\end{subequations}}
\newcommand{\bea}{\begin{eqnarray}}
\newcommand{\eea}{\end{eqnarray}}
\newcommand{\ba}{\begin{array}}
\newcommand{\ea}{\end{array}}
\newcommand{\bc}{\begin{center}}
\newcommand{\ec}{\end{center}}

\begin{document}
\preprint{IPM/P-2012/009}  
\vspace*{3mm}

\title{$\mathcal{H}$olographic $\mathcal{N}$aturalness}%

\author{Andrea Addazi}
\email{andrea.addazi@lngs.infn.it}
\affiliation{Center for Theoretical Physics, College of Physics Science and Technology, Sichuan University, 610065 Chengdu, China}
\affiliation{INFN sezione Roma {\it Tor Vergata}, I-00133 Rome, Italy}

\begin{abstract}
\noindent

The $\mathcal{H}$olographic $\mathcal{N}$aturalness ($\mathcal{HN}$) is a new paradigm towards an explanation of the Cosmological Constant (CC) and the Higgs Hierarchy (HH)
in the Universe. 
Motivated by the Holographic Principle, and inspired by the (A)dS/CFT correspondence, 
we elaborate on the possibility and on the cosmological consequences of a fundamental intrinsic disorder and temperature in {\it vacuo}. 
We postulate that the zero vacuum entropy is provided by 
a large number of quantum hair fields, the {\it hairons}. 
The quantum hairon gas in space-time induces an effective decoherence effect to the Standard Model (SM) particle sector. 
This is leading to an entropic reinterpretation of UV divergent contributions to CC and HH:
we will show that, in both the cases, the large number of re-scatterings on the hairon ensamble suppresses any radiative instabilities. 
The CC and HH problems are illusions envisaged by a conscious observer, having access on the limited amount of informations from SM tests: 
both the issues are originated from our ignorance of the hidden entropy intrinsically stored in the space-time.
The $\mathcal{HN}$ suggests to search for effective decoherence effects in particle physics observables 
such as effective CPT, Unitarity and Energy violations.
Regarding the HH, the $\mathcal{HN}$ does not introduce any new particles or interactions around the TeV-scale:
we do not expect for any signatures, at LHC and any future high energy colliders, related to the Higgs UV completion in a Wilsonian sense.

\end{abstract}

\maketitle


The observed Universe architecture appears to be organized as a hierarchical structure
from the microscopic to the cosmological scales. 
This may certainly inspire a series of philosophical if not numerological considerations
as the notorious {\it Eddington problem} as {\it why is the Universe size over the proton radius so large as 
$10^{80}$ or so?} \cite{E}. Nowadays, such a shortcomings may merely be considered 
as facts rather than magical ratios explainable by basic units of any fundamental theory of Nature. 
Very much the same can appear for the Cosmological Constant, the QCD and the electroweak scales
compared to the fundamental energetic top of the hierarchy pyramid: the Planck scale. 

A more insidious and fastidious question for the contemporary physics is the {\it why hierarchies are stable};
related to notoriously unsolved issues in the Higgs Hierarchy HH (34th digits) and the Cosmological Constant CC (123th digits) \footnote{The HH corresponds to the square of Higgs mass over the the Planck mass
as $m_{H}^{2}/M_{Pl}^{2}\sim (125\, {\rm GeV}/10^{19}\,{\rm GeV})^{2}\sim 10^{-34}$. The CC hierarchy is related to the observed  value of $\Lambda$ over $M_{Pl}^2$, which is around 
$10^{-123}$  (see \cite{Padmanabhan:2007xy} for a review on the CC problem).}. 
The anthropic principle considerations, as suggested by {\it Weinberg} in case of CC \cite{Weinberg:1987dv},
 do not really seem to unveil the dynamical 
reasons behind the hierarchical stabilization of the Universe scales. 
The concept of {\it Naturalness}, introduced by {\it t'Hooft}, 
may elegantly relate hierachies with new symmetry principles in the particle physics sector
\cite{Naturalness}. So far as we know, t'Hooft naturalness successfully works in relating the 
pion mass protection to the chiral symmetry.
These have inspired the desperate searches of new miraculous symmetries, extending the standard model sector, protecting the HH and CC. 
HH solutions suggested so far, as supersymmetry and technicolor, seem to be disfavored after LHC data.
While for the HH problem, the particle symmetry principle approach seems to be disappointingly not sustained by current LHC data, 
for the CC the situation is even worst as a completely obscure territory: 
 there is not any known realistic candidate of a symmetry transformation, extending the SM model, protecting 
the CC for 123th orders of magnitude, without any obvious logical self-contradictions.
After the many unsuccessful attempts to understand 
the HH and the CC within these paradigms, to insist on these ways may appear as a losing strategy. 
 I think that the main suspicious point of the contemporary approaches to these issues  
 is to not consider quantum gravity effects and quantum information, as point out in Ref.\cite{Addazi:2016jfq}.

Looking to hierarchies in Nature,
it is surely possible that the entropy is the {\it key} for understanding the large numbers separating the fundamental energy scales. 

As realized by {\it Bekeinstein} and {\it Hawking}, the Black Hole (BH) has a total entropy which is holographic stored in its event horizon area \cite{B,Hawking:1974sw,Hawking:1976de}.
Even if BHs are entropically holographic rather than volumetrically extensive, they are the most disordered objects in the Universe, i.e. informations cannot be stored 
in a more disorganized pattern than in BHs. 
Therefore, laws of thermodynamics can be formulated out for BH and extended to curved geometries such as the de Sitter space-time \cite{Hartle:1976tp,EQG}. 
Indeed, the very same Universe, with a Hubble radius related to the cosmological constant, contains a large amount of holographic entropy. 
This was the dawn of the Holographic principle, elaborated later on, by {\it t'Hooft} \cite{Holo1} and {\it Susskind} \cite{Holo2, Holo3} to arrive to {\it Maldacena}'s (A)dS/CFT correspondence  \cite{Maldacena:1997re,Witten:1998qj,Klebanov:1999tb}.

On the other hand, the HH and CC are commonly overlooked from a purely 
quantum field theory prospective. This leads to the commonly assumed -- and what I suggest to dub as -- {\it vacuum triviality principle}: 
the vacuum does not intrinsically store any hidden qu-bits; only the
contribution dictated by the 
Standard Model (or particle extensions) fundamental parameters may accumulate quantum information in vacuo. 
It is tacitly presumed that the vacuum does not have any intrinsic entropy if not from the SM particles/interactions.
However, this seems to point out to the opposite direction with respect to the thermodynamics laws in curved space-time,
in turn leading to the intuition that the vacuum is thermally full of information!
I think that this apparent contradiction is exactly the source of any fundamental misunderstanding of hierarchies in Nature. 
Therefore we oppose to the vacuum triviality and we will go ahead departing from common starting points considered in any current CC and HH considerations. 

If the vacuum information storage is holographically scaling,
then its intrinsic entropy is 
$$S_{in\,vacuo}\sim A/L_{Pl}^{2}\,,$$
where $A$ is the holographic area, $L_{Pl}$ the Planck length.
In a de Sitter space-time, the entropy is related to the Hubble area and the cosmological constant
as 
$$S_{de\, Sitter}\sim r_{\Lambda}^{2}/L_{Pl}^{2}\, ,$$
where $r_{\Lambda}$ is the Hubble radius of the Universe. 
 Therefore, the Universe has a fundamental 
temperature of $$T\sim \sqrt{\Lambda}\, .$$
 Now, this temperature is appearing out as a warning:
for a so high entropy of the Universe, the dS-space-time entropy is 
$$S_{\Lambda}\sim 10^{123}\, : $$
 the Standard Model degrees of freedom are not enough for accounting 
it. As pointed out by {\it Penrose}, the probabilistic configuration space covered by the all baryons in the Universe today 
is practically zero compared to the whole possible configurations allowed \cite{Penrose}. 
The baryon amount is of the order of the Eddington number $N_{E}\sim 10^{80}$.
Every baryons count as an entropic contribution of $\sim 10^8$ with a Cosmic Microwave Background temperature of $T_{CMB}\simeq 2.7\,{\rm K}$.
Therefore the total amount of radiation entropy in the Universe is 
$$S_{U}\sim 10^{88}\, ,$$ 
that is much below to the the entropy amount predicted by holography in vacuo. 
Intriguingly, the maximal entropy state corresponding to an Eddington number of baryons is 
exactly of the order of $10^{123}$. 
An enormous landscape of configurations in the Universe probability space, predicted by holography,
$$\Omega_{U}=e^{S_{U}}\sim 10^{10^{123}}\, >>> \Omega_{B}=10^{10^{88}}\, ,$$
is not covered by ordinary fields. 
On the other hand, the radiation entropy cannot be related to the fundamental temperature 
of the de Sitter Universe in presence of a cosmological constant, 
since they cannot provide for any Universe acceleration mechanism. 
As regards the SM vacuum state, it can only encode for a very limited amount of informations from the particle physics parameters.
This is insightfully suggesting that the SM vacuum is only a limited sub-structure of the cosmological vacuum state!
It cannot provide more than just a sub-contribution to the vacuum repulsion sourcing for the Universe acceleration.
For avoiding any information lost paradox, we can postulate that the vacuum is enriched by a large number of {\it quantum hairs},
encoding for missing (qu)bits.
The quantum hairs $h_{1,...,n}$ encode the missing information, filling the maximal probability configuration
$$\Omega(h_{1},...,h_{n})\sim 10^{10^{123}}\, .$$

The presence of quantum hairs was suggested by many authors as a way out to the information lost in black holes
as well as the classical no-hair theorem 
\cite{V1,Coleman:1991ku,Preskill:1992tc,Giddings:1993de,Dvali:2012rt,V2}.

\begin{figure}[ht]
\centerline{ \includegraphics [width=1\columnwidth]{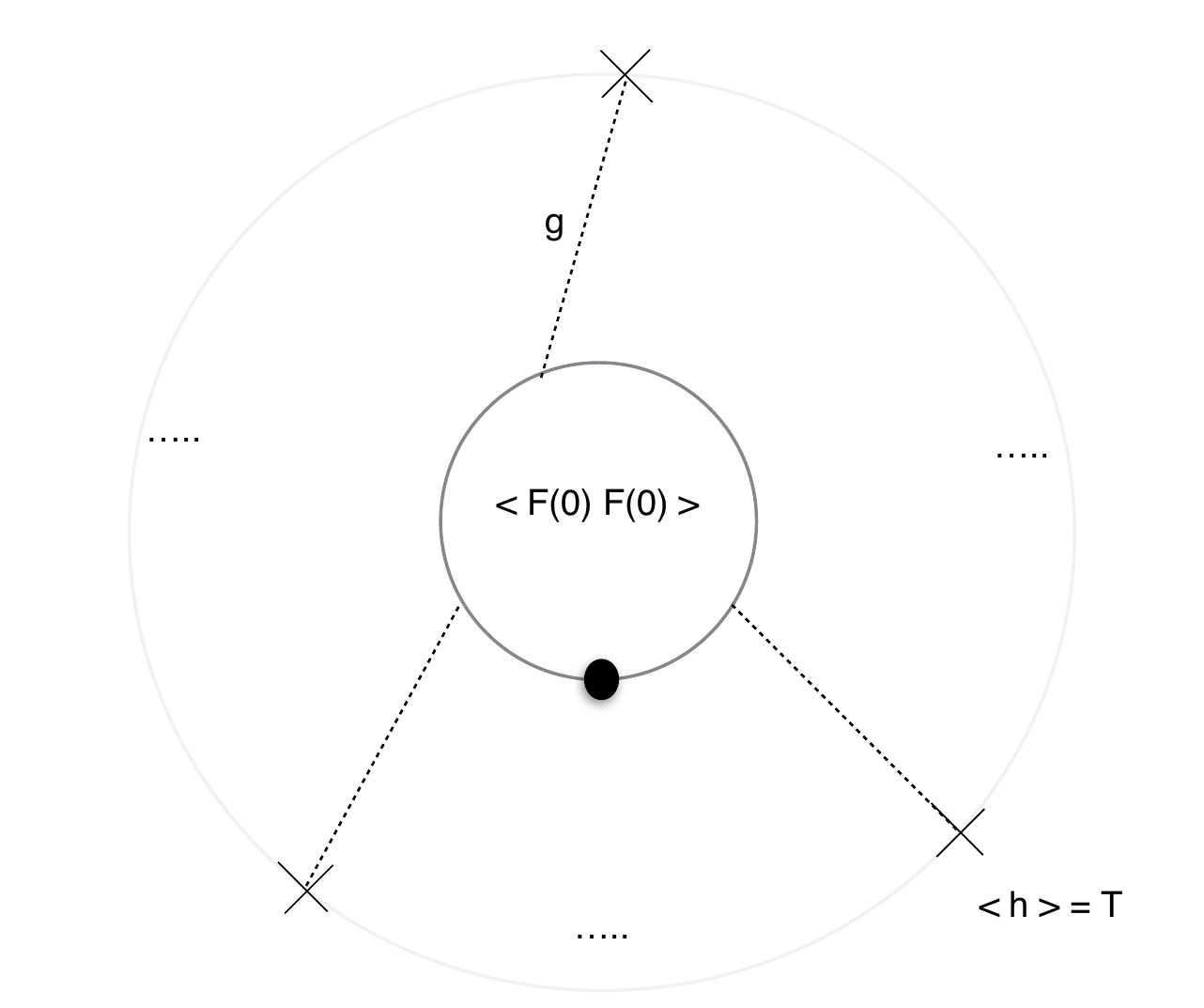}}
\caption{The standard model vacuum bubble diagrams for any fields $F$, corresponding to $\langle F(0) F(0)\rangle$, have N-graviton insertions from hairon background fields, with a thermal expectation value of $\langle h \rangle=T$.}
\end{figure}

\begin{figure}[ht]
\centerline{ \includegraphics [width=0.7\columnwidth]{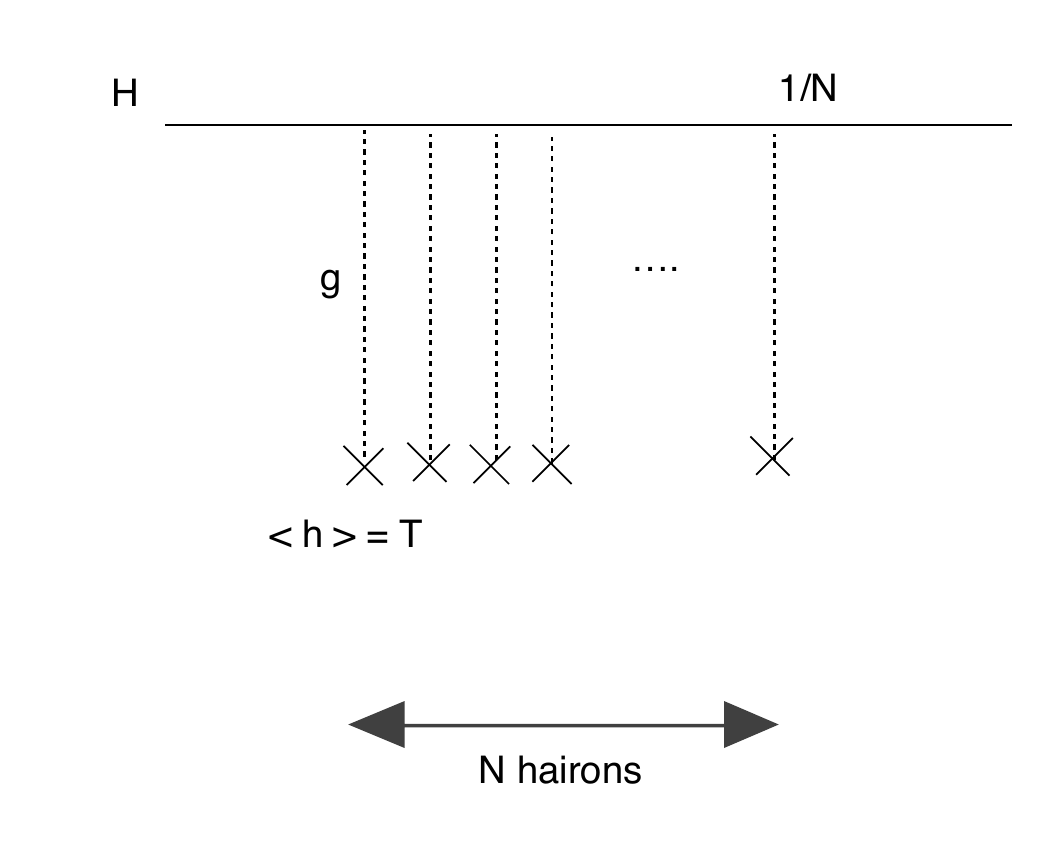}}
\caption{The Higgs propagator has a large N-number of insertions from hairons, in turn with a thermal expectation value $\langle h \rangle=T$. The single hairon is weakly coupled with the Higgs through a graviton as $\alpha_{G} \sim 1/N$. 
However, the Higgs is collectively strongly coupled with the hairon ensamble in a scrambling time $t\sim \sqrt{N}t_{Pl}\, \log\, N$, which two digits close to the electroweak time.  }
\end{figure}

The holographic entropy quadratically scales with the observation length
and this will be the main point towards all our considerations on hierarchies.
We claim that hierarchies are holographically ordered, as a new naturalness paradigm that we dub the $\mathcal{H}$olographic $\mathcal{N}$aturalness $(\mathcal{H}\mathcal{N})$.
This is also motivated by the dS/CFT paradigm \cite{Strominger:2001pn,Bousso:2001mw}.
In our picture, a quantum gas of hairs, stored in the space-time volume with a constant number density, is envisaged.
The quantum hair gas has an averaged kinetic energy 
related to a temperature in the vacuum state. 
This picture promotes hairs to dynamical fields, dubbed {\it hairons}.
In a Hubble radius, the hairon gas temperature is 
$T\sim \sqrt{\Lambda}$.
The dS entropy is provided by the hairon counting as 
$$S\sim N\sim M_{Pl}^{2}/\Lambda$$
and 
$$T\sim \sqrt{\Lambda}\sim M_{Pl}/\sqrt{N}\, .$$
The hairon gas is not an ideal one, its pressure does not diluite 
with the volume, since hairons are stored in any fundamental volume cell of space-time;
their equation of state has a dark energy like form 
$P=-\rho\sim -M_{Pl}^{2}T^{2}$, with $w=-1$. Indeed, the vacuum energy, which thermodynamically is commonly assumed 
as the internal energy, is not the only possible source of the Universe expansion.
In our case the Universe expands because of its entropic energy repulsion\footnote{In a broad sense this may be considered in analogy with the Casimir effect, where the plates 
attract each others from both the vacuum energy pressure as well as the thermal one.}.

Within this picture, any Standard model correlators have to be thermally averaged inside the 
vacuum state $|T\rangle\simeq |N\rangle$. This is leading to an effective decoherence effect 
propagating into the SM sector. 

 Let us consider the usual radiative UV contributions to the vacuum energy density:
the energy-momentum tensor vacuum expectation value receives large contribution as 
$$\langle T_{\mu\nu}\rangle=-\rho_{UV}g_{\mu\nu}\, ,$$
where the $\rho_{UV}$ is proportional to UV divergent correlators. 
The main quantum contributions to the vacuum energy density are provided by Feynman bubble diagrams
In SM, the bubble diagram corresponds to 
$$\langle F^{2}(x) \rangle\equiv \langle 0|F^{2}(x)|0\rangle\, ,$$
 where $|0\rangle$ is the empty vacuum state and $F$ are any possible SM fields;
but in our case we must consider such a correlator in the thermal vacuum state 
as $\langle T|F^{2}(x)|T\rangle$ (see Fig.1). 
This is offering an insightful reinterpretation of the CC quantum corrections. Now, the computation of this diagram is
in thermal field theory and it is now leading to quartic contributions as 
$$\Delta \rho_{\Lambda}\sim (n_{B}-n_{F})T^{4\, }\, .$$
However, $T\simeq \sqrt{\Lambda}$ and therefore we obtain natural corrections! Sending $\Lambda\rightarrow 0$ 
all higher thermal corrections flow to zero as well. 
The UV energies, flowing in propagators within SM computations on a non-thermal vacuum, 
are now thermalized by the insertion of many re-scatterings of SM particles on the thermal hairon background.
In other words, UV divergences in the SM subsector of the vacuum state are washed out.
The UV SM radiative corrections are entropically suppressed in the thermal bath:  
$$\langle T|F^{2}(x)|T\rangle=\langle T|0\rangle \langle 0|F^{2}(x)|0\rangle \langle0|T\rangle\, ,$$
where 
$$\langle T|0\rangle=\langle 0|T\rangle^{*}$$ are transition amplitudes suppressed by the thermal entropy content 
$$\langle T|0\rangle=e^{-S(T)}=\frac{1}{\Omega(T)}\, .$$ 
Therefore, any UV divergences, even Planckian colossal contributions, are 
dressed as a suppression factor 
$$\sim e^{-S} M_{Pl}^{4}\sim e^{-10^{123}}10^{123}\Lambda\rightarrow {\it forget{-}about{-}it}\, !$$
\footnote{At this point it is our duty to make a comment: 
it is hard to me to think about a completely thermalized vacuum state.
Within this mind, the thermal vacuum state is consider as a good approximation 
of averaging on a large number of hairons, as $\Lambda\sim  T^2=M_{Pl}^2/N$. 
In the limit of $N>>1$, deviations to the entropy from perfect thermality flow down, as inverse powers of $N$, to negligible contributions. }
The UV quantum fluctuations would lead the system into a more ordered sub-region of the probability configurations 
and, therefore, they are highly suppressed as the configuration volume factor $\Omega(T)$. 

It is worth to remark that the thermal stabilization mechanism of the CC can efficiently work if 
all the SM wave functions are fully entangled with the hairon wave function. 
The characteristic time for a full entanglement may naively appears 
as extremely long, if considering a so large number of hairon fields.
However, we should not forget that holographic systems have a peculiarly fast scrambling time  \cite{Hayden:2007cs,Sekino:2008he,Shenker:2013pqa},
as 
\begin{equation}
\label{scra}
\tau\sim \Lambda^{-1/2}\, \log \, \frac{M_{Pl}^{2}}{\Lambda}\, . 
\end{equation}
In the N-hairon picture, this corresponds to 
\begin{equation}
\label{scra2}
\tau \sim \sqrt{N}\, t_{Pl}\, \log \, N\, . 
\end{equation}
The scrambling time is long but only two digits than the Hubble time:
\begin{equation}
\label{scra2}
\frac{\tau}{t_{H}} \sim \log \, N\, \simeq 280\, . 
\end{equation}
This reduces the fine-tuning level from $10^{123}$ to only two digits.

\vspace{0.1cm}

Let  us now consider the HH problem.
The Higgs has an electroweak order life-time; is it enough for its efficient thermalization in {\it vacuo}?
Since the harions are hidden to any colliders direct production, we necessary have to assume that they are weakly coupled 
with the Higgs boson. Therefore, we assume that the only interaction portal among the Higgs and hairons is gravity. 

Let us assume that the Higgs has a bare mass around the e.w. scale. Then its wave function explores an electroweak space-time volume
and, therefore, according to the holographic scaling, an entropy of 
$$S\sim N=M_{Pl}^{2}/m_{H}^{2}\sim 10^{34}\, .$$
Every gravitational Higgs-hairon coupling is 
$$\alpha_{G}(E)=E^2/M_{Pl}^{2}\sim N^{-1}\, , $$
related to a single hit collision time $\tau \sim (\alpha_{G}^{-1/2})t_{Pl}\sim \sqrt{N} t_{Pl}$.
Because of the holographic criticality state, soon after the first hit time,
the full entanglement time has a characteristic scaling as 
\begin{equation}
\label{scram}
\tau \sim m_{H}^{-1}\, \log \, \frac{M_{Pl}^2}{m_{H}^2}\sim \sqrt{N}t_{Pl}\,\log\, N\, ,
\end{equation}
where $m_{H}$ is the Higgs mass. This is the same law introduced for CC above. 
Therefore, the Higgs is efficiently thermalized in a time that, plugging inside
Eq.\ref{scram} the $N\sim 10^{34}$, corresponds to around $100\, t_{H}$,
where $t_{H}$ is the Higgs life-time. Therefore the fine tuning is now of two digits rather than 34th.
 Indeed, after the scrambling time transient, hairons efficiently transfer to the Higgs their average thermal energy 
$$T=M_{Pl}/\sqrt{N}\sim m_{H}$$ 
within the electroweak space-time volume.

The Higgs boson propagator has a large number of re-scattering contributions from the thermal hair background (see Fig.2).
Because of that, also the Higgs loop corrections are thermally averaged as in thermal field theory.
The Higgs mass corrections are proportional to the average thermal bath kinetic energy $T$;
the one thermal loop corrections to the Higgs are as the leading terms 
$$(c_{B}n_{B}-c_{F}n_{F})T^{2}\sim \frac{1}{N}(c_{B}n_{B}-c_{F}n_{F}) M_{Pl}^{2}$$
 plus subleading thermal corrections, 
where one consider all SM fermions and bosons couplings $c_{B,F}$ and their numbers. 

If the Higgs boson has a bare mass close to the electroweak scale, then 
it is stabilized as a thermal collective phenomena inside the electroweak world. 
Inside, the e.w. volume, the holographic scaling predicts a temperature $T$ of the very same order of the electroweak scale,
in turn related to the Higgs mass. Therefore, as for the CC, the Higgs is Holographically Natural. 

From the $N$-prospective, the thermal suppression 
factors $\langle 0|T\rangle={\rm Exp}\{-S\}$ approximately correspond to $\langle 0|N\rangle={\rm Exp}\{-N\}$, 
and therefore CC and HH suppressions are 
$$\langle N| F^2(x)|N\rangle=e^{-2N}\langle 0| F^2(x)|0\rangle$$
 and 
$$\langle N| F^{\dagger}(x)F(y)|N\rangle=e^{-2N}\langle 0| F^{\dagger}(x)F(y)|0\rangle\, .$$

These {\it full-empty} vacuum transitions also eliminate all tadpole diagrams in the standard model, probabilistically exponentially disfavored in every Feynman diagrams considered. 
It is well known that in standard model, we just use to forget about tadpole diagrams,
but really without any theoretical justification. Here, we naturally explain why tadpoles are not relevant. The vacuum selects 
among all possible Feynman loops, as an information razor principle. 
Formally, any tadpole 
$$\langle N| F^{2}(x)F^{\dagger}(x)F(y)|N\rangle$$
 is suppressed as 
 ${\rm Exp}\{-2N\}\,$.

\vspace{0.3cm}

{\it Holographic Decoherence}. Let us consider the Universe wave function of the Standard Model (or any extensions) vacuum state as an undetermined superposition of all possible cosmological constant states as 
\begin{equation}
\label{Psi}
|\Psi \rangle=\sum_{N}\Psi_{N}|N\rangle\, , 
\end{equation}
with energy eigenvalues of $|N\rangle$ equal to $E_{N}=\sqrt{\Lambda}_{N}=M_{Pl}/\sqrt{N}$ and where
$\Psi_{N}=\langle N|\Psi\rangle$ are the probability amplitudes that the Universe wave function 
is in a certain $|N\rangle$ state. 
From the standard arguments prospective, one ignores energy discretization and  
\begin{equation}
\label{Psi}
|\Psi \rangle=\int d\Lambda |\Lambda\rangle \langle \Lambda|\Psi\rangle\, , 
\end{equation}
and continuous spectrum $\Lambda$. 
These can be thought as eigenvectors of the Wheeler de Witt equation \cite{dW}
as well as reformulated in a path integral approach \cite{Hartle}. 

A posteriori, we know that $N$ must be large as $N\sim M_{Pl}^{2}/\Lambda\sim 10^{123}$.
Rephrased in this way, one would start to rise intriguing questions. Why and How would the wave function collapse to 
a specific and so low cosmological constant state as the one observed? 

This is  leading back to the measurement problem in quantum mechanics. 
Here, we elaborate on the quantum decoherence picture.
Indeed, we find difficult to immagine  that any specific humanoid observer may reduce the vacuum wave function.
For example, this would lead to a series of serious confusions regarding causality and locality. 
Avoiding this ambiguous territory, the only way seems to be on the objective reduction:
the Standard Model vacuum state must interfere with other hidden states $|H\rangle$ which lead to the effective decoherence 
of it. The hidden state contains the quantum hairs stored in space-time as $|H\rangle=|h_1,h_2,....,h_M\rangle$ where $h_{i}$ are hair quanta.  
In other words, the SM vacuum state, namely $|\Psi\rangle$ is a sub-wave function of Universe, ignoring the zero entropy state
which can be describe by  $|H\rangle$. The complete wave function 
is $|\Phi\rangle=|\Psi \rangle |H\rangle\equiv |\Psi \rangle \otimes |H\rangle$. 
We can introduce the Universe density-matrix as 
\begin{equation}
\label{rho}
\rho=|\Phi \rangle \langle \Phi|=|\Psi \rangle |H\rangle  \langle H|\langle \Psi|\, . 
\end{equation}

Now, when the sub-system $|\Psi\rangle$ fully interacts with the environmental state
$|H\rangle$, then this corresponds to an effective tracing out, as 
\begin{equation}
\label{rho}
\rho_{\Psi}={\rm Tr}_{H}(\rho)=\sum_{H}|\Psi\rangle \langle \Psi| |\langle H| H\rangle|^{2}=|\Psi\rangle \langle \Psi|\, 
\end{equation}

The gravitational interaction between the the SM system and the hidden environment is unavoidable,
as a gravitational-mediated decoherence effect. As we saw before, the thermalization 
is efficient enough for explaining any hierarchies.

After the full interaction or thermalization, the sub-system $|\Psi\rangle$ would collapse to a specific eigenstate $|N\rangle$:
\begin{equation}
\label{deco}
|\Psi\rangle |H\rangle \rightarrow |N\rangle |H\rangle \, .
\end{equation}

From the non-interacting to the interacting phase there would be a time transient $\Delta t$.
Therefore we have two different density-matrices for the non-interacting and interacting phases:
we dub $\rho_{0}$ the full density-matrix of the non-interaction stage and $\rho_{\Psi}$ as the one tracing out the hidden environment. 
Let us consider the generic probability of a transition $|\Psi\rangle\rightarrow |\zeta\rangle$ in the non-interaction epoch: 
$$P_{\Psi}(\Psi \rightarrow \zeta)=\langle\zeta|\rho_{0}|\zeta \rangle=\langle \zeta|\Psi\rangle \langle\Psi|\zeta \rangle$$
\begin{equation}
\label{sumN}
=\sum_{N}|\Psi_{N}^{*}\zeta_{N}|^{2}+\sum_{N\neq M}\Psi_{N}^{*}\Psi_{M}\zeta_{M}^{*}\zeta_{N}\, , 
\end{equation}
where $|H\rangle$ and $\langle H|$ are just contributing as the identity here and we expressed the state $|\zeta\rangle$ 
as a linear combination of the basis $|N\rangle$ with amplitudes $\zeta_{N}=\langle N|\zeta\rangle$. 
When we use $\langle H|H\rangle=1$, we are assuming the hidden quantum states as a complete set, we do not introduce any information incompleteness here,
probabilities are conserved. 
The last contribution in Eq.\ref{sumN} corresponds to the quantum interference. 
Now, let us consider the same problem in the interacting phase, 
where the density-matrix has the trace-out form 
\begin{equation}
\label{densi}
\rho_{\Psi}=\sum_{N}|\Psi_{N}|^{2}|N\rangle  \langle N|\, .
\end{equation}
Then the transition probability has the form 
\begin{equation}
\label{densid}
P(\Psi\rightarrow \zeta)=\sum_{N,M}|\Psi_{N}|^{2}|\zeta_{M}|^{2}\delta_{MN}=\sum_{N}|\Psi^{*}_{N}\zeta_{M}|^{2}\, , 
\end{equation}
where interference terms $\sum_{M\neq N}\Psi_{N}^{*}\Psi_{M}\zeta_{M}^{*}\zeta_{B}$ are decoherently elided. 

If $|\zeta\rangle$ just coincides with an eigenstate $|N\rangle$ (or in the continuum limit $|\Lambda\rangle$)
then the $|\Psi\rangle$ must inevitably collapse to the very same state $|N\rangle$ with $100\%$ probability. 

Let us now consider that the the environment does not provide for a single interaction with the sub-system
as a sort of single measure. After the $\Psi$-function collapse, the environment will continue to interact with it,
thermalizing it. The $|H\rangle$ is in turn expressed as a basis of energy eigenvectors, which, in general can 
be different than the $\Psi$-basis. The Hilbert (or, in QFT, the Fock) spaces can be $\mathcal{H}_{1}\otimes \mathcal{H}_{2}$
and $|H\rangle \otimes |\Psi\rangle$ an entangled state of the two. 

Now, if $|H\rangle$ is just on a eigenstate $|N\rangle$, then $\rho_{\Psi}$ would just trivialize to the identity matrix. 

The SM zero state $|\Psi\rangle=|0\rangle$ can be considered as 
the sum all over SM oscillator vacuum energies $\sum_{SM} \hbar \omega$. 
This vacuum energy density tends to be UV divergent, leading to the CC problem:
it seems that SM vacuum energy does not spontaneously sit down to the tiny as the observed. 
However, after thermalization, this sub-state is fully entangled with the $|H\rangle$. 
Inspired by the Holographic principle, a dS space-time has a thermal state 
$|H\rangle=|N\rangle$ that has a vacuum thermal energy eigenstate $T\sim \sqrt{\Lambda}=M_{Pl}^{2}/N$. 
Therefore, if the subsystem $|\Psi\rangle$ is fully interacting with $|\Lambda\rangle=|N\rangle$,
the decoherence will only allow to it to collapse to $|\Lambda\rangle$, or at least to a degenerate set 
of common vacuum states with respect to other observables $O$ as $|\Lambda, O\rangle$. 
Indeed the SM oscillator vacuum energy can be thought as associated to 
a series of Gaussian zero-wave functions that, inside the thermal bath, are dissipated out 
after a certain time transient.

\vspace{0.1cm}

{\it On the possible origin of hairons.} When we introduced the concept of hairons, 
we introduced it as a generic paradigm. An interesting issue is how the 
hairons are originated from. In other words, how can we interpreted the presence of these
new fundamental degrees of freedom stored in space-time?
One possibility is that hairons emerged as Goldstone bosons of a spontaneous breaking 
of a large or even infinite global symmetry, namely a $\mathcal{M}_{\infty}$-symmetry. 
Recently, many authors have suggested that the information recast in black holes 
is explained from a new infinite global symmetry, such as BMS \cite{Strominger:2013jfa,Hawking:2016msc} \footnote{Recently, a new time dependent super-translation symmetry was suggested in Ref. 
\cite{Chiang:2020lem}.} and Kac-Moody algebras
\cite{Ellis:1991qn,Ellis:2016atb,Addazi:2017xur}.
The role of these infinite symmetries were also discussed in vacuum and in de Sitter cases \cite{Addazi:2017xur,Hamada:2017gdg}. 
It is conceivable that the many quantum hairs stored in {\it vacuo}, envisaged by various authors,
may simply coexist as a jeopardized state. To distinguish the observable effects of a certain class to another may appear as desperately impossible.
Therefore, we prefer to refer to all these possibilities as the only classification name of hairons.

\vspace{0.5cm}

{\it On the recent discussions on de Sitter existence}. Recently many authors discussed 
possible arguments against the existence of a pure de Sitter space-time, motivated by the
Swampland conjecture \cite{Obied:2018sgi}.
Since in our paper we largely discussed about holography and de Sitter, 
it would be our duty to make a comment on it. 
First of all, we do not think the our idea can restrict the possibility of 
a dynamical dark energy, as quintessential or modified gravity inspired scenarios. 
For example, even if the hairon get expectation value is $\langle h\rangle\sim T$, it is certainly possible 
that they have an effective potential rendering their dynamics highly non-linear and leading to a secularly time variation of the temperature.
Outside any specific frameworks, we cannot preclude both dynamical or non-dynamical cases.
Both scenarios are afflicted by hierarchy problems and in our case we suggest a holographic solution. 
Then, we highly suspect that our scenario does not favor for any eternal de Sitter vacua,
following some analogous thinking pathways inspired by information scrambling in Bose-Einstein gravity 
\cite{Dvali:2018fqu,Dvali:2018jhn} as well as the presence of S-brane or CFT-Liuoville instabilities in dS/CFT \cite{Gutperle:2002ai,Sen:2002nu} and, finally, by tunneling processes of black hole pairs. 
In all these cases, the instabilities are expected to be longer than the Universe age \cite{EQG}. 

\vspace{0.1cm}

{\it Phenomenology of Holographic Naturalness.}
The Hidden sector simulates an information lost in the SM sector 
for an observer who does not have any direct access to it. 
The probabilities are not conserved in the SM sector,
unitarity is lost, CPT is violated, energy is not conserved. But only in the accessible sub-sector 
and not in the complete theory. 
Therefore, $\mathcal{H}$olographic $\mathcal{N}$aturalness suggests to search for new physics from
the apparent lost of coherence in SM observables. 
There are many interesting channels for testing it. 
Certainly, high energy collider physics as from LHC data do not appear to be the best strategy for searching for 
$\mathcal{H}\mathcal{N}$ footprints. From this point of view, $\mathcal{H}\mathcal{N}$ appears fundamentally different than other suggestions such as 
TeV-supersymmetry and composite Higgs models. 

Possible apparent decoherence phenomena can be searched in several
different sectors, including Kaon-AntiKaon transitions \cite{E,Mavromatos:2018rds},
Neutron-Antineutron physics \cite{Babu:2016rwa,Addazi:2015oba}, high precision interferometers \cite{Verlinde:2019xfb},
entangled systems \cite{E1,E2},
velocity dispersions in {\it vacuo} in Very High Energy Cosmic neutrinos \cite{Ellis:2011uk,Amelino-Camelia:2016ohi,Zhang:2018otj}
and many other possible signatures.   
In the case that hairons may have a total angular momenta different than zero,
in principle they can induce apparent energy-dependent Pauli Violating transitions 
that can be searched in Underground experiments \cite{Addazi:2017bbg}. 
For example, it is possible that certain hairons are organized as vortices in {\it vacuo}
and a SM particle scattering on them may exchange a tiny angular momenta. 

\vspace{0.2cm}

{\it More about the measurement problem. The Universe Brain.} 
The $|\Phi\rangle=|H\rangle\otimes |\Psi\rangle$ state is entangled by the gravitational interactions.
The thermalization consists in a re-organization of the the qu-bits contained in the entangled state
space $|\Phi\rangle$.
The effective decoherence effect is nothing but a re-elaboration of qu-bits from an initial state
$$|\Phi\rangle=|\phi_{1},..,\phi_{N}\rangle \equiv |h_{1},...,h_{n}\rangle \otimes |q_{1},..,q_{m}\rangle$$
to another entangled state
$$|\Phi'\rangle=|\phi_{1}',..,\phi_{N}'\rangle$$
where 
$$|\Phi'\rangle = U |\Phi\rangle\, $$
and the unitary transformation $U$ can be considered as a quantum computation elaborating in-put information into 
out-put qu-bits. 
In the SM $+$ Hidden states, the SM qu-bits are much less than the Hidden ones. 
Therefore, the computation 
$$|\Phi\rangle=|\Psi\rangle |H\rangle \rightarrow |N\rangle |H'\rangle\simeq |N\rangle |H\rangle$$
is appearing as an effective decoherence since, while the SM superpositions of $|N\rangle$ would collapse to one eigenstate selected by the
Hidden one, the $|H\rangle$ Hilbert space is practically un-changed. 

Therefore, we can fictitiously think the $|\Psi\rangle$ as the quantum microscopic system and the $|H\rangle$ as 
a Macroscopic apparatus or even as a quantum computer. 
The $|H\rangle$ effectively measures the $|\Psi\rangle$ sub-state, collapsing it. 
Within the transition \ 
$$\langle \Phi'|\Phi\rangle=\langle \Phi|U|\Phi\rangle=\langle \Psi|\langle H|H'\rangle |N\rangle\simeq \langle \Psi|N\rangle\, ,$$
one would expect an out-of-equilibrium relaxation transient where information scrambling 
may be envisaged as a quantum chaos effect \footnote{If $|H\rangle=|N,O_{1},...,O_{N}\rangle$, then, from tracing out the $|H\rangle$ we also sum on all over $N$, obtaining 
$\langle \Phi'|\Phi\rangle\simeq 1$.}.

Now, as pointed out by {\it Dvali}, there are intriguing analogies among the space-time holographic information storage 
and neural networks \cite{Dvali:2017ktv,Dvali:2018vvx}. 
In Brain networks and BH, the memory storage is highly enhanced and both exhibiting an area law.
The $|H\rangle$ state works as a Brain network, elaborating the in-put informations from the $|\Psi\rangle$.
In other words, the de Sitter space-time works as a {\it Universe Brain} elaborating any information 
entering in contact with.

\noindent

{\it Conclusions.} We have discussed a new paradigm for reinterpreting the Naturalness 
of the Cosmological Constant (CC) and the Higgs Hierarchy (HH) in light of the Holographic principle.
We moved from a simple but, we believe, deep consideration: 
the Holographic de Sitter has an entropy $S_{\Lambda}\sim M_{Pl}^{2}/\Lambda\sim 10^{123}$ 
enormously larger than the SM information contained in the Universe. 
This sounds as a hint that the vacuum structure is much richer 
than expected, with a large amount of hidden qu-bits stored in every Planckian volume of space-time. 
The missing information can be accounted by dynamical quantum gravity hairs, the hairons. 
A series of powerful relations follows as
$$S_{\Lambda}\sim M_{Pl}^{2}/\Lambda\sim N \sim M_{Pl}^{2}/T^{2}\, .$$

From such correspondences,
we obtain strong unexpected implications on CC and HH issues. 
First of all, if $\Lambda\sim T^{2}\sim N^{-1}M_{Pl}^{2}$,
then:

\vspace{0.1cm}

 i) the CC is a thermal effect;  the temperature is the average kinetic energy of 
a quantum gas of hairons;

\vspace{0.1cm}

 ii) any usual SM quantum corrections are thermally averaged 
and they contribute as powers proportional to the CC temperature itself $T\sim \sqrt{\Lambda}$, at any loop-orders;

\vspace{0.1cm}

iii) the Higgs propagator is thermalized and the loop corrections scale as the temperature inside the 
electroweak space-time volume (probed by the Higgs). Quantum loop corrections are thermally suppressed to powers of $T\sim m_{H}$, similarly to the CC case.

\vspace{0.1cm}

Therefore, the CC and HH are natural, since the whole tower of their UV quantum corrections is replaced by thermal ones
which are proportional to the CC and HH bare parameters; i.e. all corrections flow to zero when $m_{H}, \Lambda\rightarrow 0$.
This is compatible with t'Hooft naturalness principle, even if in a different and unexpected way since
there is not any new symmetry extending the particle physics sector. This is a new form of naturalness,
the ${\mathcal{H}}$olographic ${\mathcal{N}}$aturalness. These phenomena may be interpreted as 
a collective stabilization effect provided by the environment and entropic effects. 

We have also discussed the thermalization of any SM vacuum instabilities as a decoherence effect.
A conscious observer only accessing to SM informations, not having any possibility to probe the 
quantum hairs, would be puzzled by hierarchy instabilities. However, these problems are illusions: 
they are effects of the observer ignorance of the whole hidden information. 

The Hidden hairon gas acts as an apparatus or an observer measuring and collapsing the SM wave function to 
a cosmological constant, relaxed down to the Universe intrinsic temperature. This is an effective decoherence 
effect originated from the SM vacuum entanglement with the the Hidden state, through graviton mediations.  
Information processing inside the Hidden system is holographically scaling as in neural networks \cite{Dvali:2017ktv,Dvali:2018vvx}. 
This suggests that the Hidden Universe is acting as a Universe Brain elaborating the SM informations. 
Within this picture, any CC and HH quantum instabilities are elaborated in the Universe Brain and exponentially relaxed down. 
In this sense, the CC and HH relaxations are re-interpreted as computations of the Universe Brain. 

The ${\mathcal{H}}$${\mathcal{N}}$ paradigm suggests a rich phenomenological exploration 
in searching for apparent decoherence, CPT Violations, Energy Violations, Unitarity Violations, Lorentz Violations and even Pauli Exclusion Principle Violations
in any SM observables. 

On the other hand, we remark, once again, that  ${\mathcal{H}}$${\mathcal{N}}$ on the HH does not predict any new particles and interactions
 in high energy colliders as, and beyond, LHC.
 In this sense, the ${\mathcal{H}}$${\mathcal{N}}$ paradigm 
leads to a vision of the HH not far from the
 {\it Dvali}'s \cite{Dvali:2019mhn} and {\it Senjanovic}'s \cite{Senjanovic:2020pnq} recent discussions:
Higgs naturalness does not necessary have one unique bridge to the TeV-frontier; 
new forms of naturalness may emerge out without introducing any new Higgs UV completion in a Wilsonian sense. 
Indeed, the ${\mathcal{H}}$${\mathcal{N}}$ provides an example of {\it Collective Completion}, eliminating UV divergences without 
introducing any new heavy modes in the theory. 

The Holographic Principle appears to have a continuous and irresistible 
predictive power, now unexpectedly extending its domain to the CC and HH problems. 
The deepest mechanisms
behind laws and order in Nature may be unveiled from the ${\mathcal{H}}$${\mathcal{N}}$ prospective. 

${\mathcal{H}}$${\mathcal{N}}$ has the potentiality to be a revolutionary paradigm 
for our understanding of why the hierarchy pyramid, from the Universe radius to the Planck length,
is not unstable, and not falling down to the quantum gravity domain, as a castle made of sand. 

\vspace{0.5cm}

\vspace{0.5cm}

\noindent

%


\onecolumngrid


\twocolumngrid

\end{document}